\documentclass{article}
\usepackage{spconf,amsmath,graphicx}
\usepackage{color}
\usepackage{caption}
\usepackage{subcaption}
\usepackage{amsfonts}
\usepackage{bm}
\usepackage{blindtext}
\usepackage{svg}
\usepackage{multirow}
\usepackage{placeins}

\def\blfootnote{\xdef\@thefnmark{}\@footnotetext}

\graphicspath{{figures/}}

\def\comment #1 {\textcolor{red}{#1}}
\def\rts{RT$_{60}$}

\title{Attention Is All you Need For Blind Room Volume Estimation}

\name{$
\begin{array}{ccc}
	\mbox{Chunxi Wang$^{1}$, Maoshen Jia$^{1}$, Meiran Li$^{1}$, Changchun Bao$^{1}$, Wenyu Jin$^{2}$}.
\end{array}
$}

\address{
$^{1}$  Speech and Audio Signal Processing Laboratory, Faculty of Information Technology, 
\\Beijing University of Technology, Beijing, China\\
$^{2}$ AcousticDSP Consulting LLC, St Paul, MN, United States
} 

\begin{document}
\ninept
\def\baselinestretch{.903}\let\normalsize\small\normalsize
\maketitle
\begin{abstract}
In recent years, dynamic parameterization of acoustic environments has raised increasing attention in the field of audio processing. One of the key parameters that characterize the local room acoustics in isolation from orientation and directivity of sources and receivers is the geometric room volume. Convolutional neural networks (CNNs) have been widely selected as the main models for conducting blind room acoustic parameter estimation, which aims to learn a direct mapping from audio spectrograms to corresponding labels. With the recent trend of self-attention mechanisms, this paper introduces a purely attention-based model to blindly estimate room volumes based on single-channel noisy speech signals. We demonstrate the feasibility of eliminating the reliance on CNNs for this task and the proposed Transformer architecture takes Gammatone magnitude spectral coefficients and phase spectrograms as inputs. To enhance the model performance given the task-specific dataset, cross-modality transfer learning is also applied. Experimental results demonstrate that the proposed model outperforms traditional CNN models across a wide range of real-world acoustics spaces, especially with the help of a dedicated pretraining and data augmentation schemes.
{\let\thefootnote\relax\footnote{{\noindent This work was supported by the National Natural Science Foundation of China under Grant No. 61971015 and Beijing Natural Science Foundation (No.L233032, L223033).}}}
\end{abstract}


\section{Introduction}\label{sec:intro}
Dynamic parameterization of acoustic environments that users evolve has become an emerging topic in recent years. Parameters that characterize local rooms or other acoustic spaces can be used to model or design audio filters for various applications, e.g. speech dereverberation for automatic speech recognition (ASR) and voice communication \cite{Zhang18,Wu17}, spatial sound systems with room equalization \cite{Cecchi18,Jin15} etc. In particular, for the proper realization of audio augmented reality (AAR), virtual acoustic objects are required to be seamlessly integrated into the real environment, which makes a good match between acoustical properties of virtual elements and the local space a necessity \cite{Neidhardt22}.

Conventionally, measured room impulse responses (RIRs) can be used to directly derive room parameters such as reverberation time (\rts) and direct-to-reverberant ratio (DRR). Another position-independent parameter, which has been proposed as a key part of the so-called “reverberation fingerprint” of a room, is the geometric room volume $V$.
Under ideal diffuse sound field assumptions, the relation between these parameters is given by the widely known Sabine’s equation \cite{Kuttruff16}:
\begin{equation}
    RT_{60}(b) \approx 0.16  \frac{V}{\alpha(b) \cdot S},
\end{equation}
where $S$ denotes the total area of the room’s surfaces and $\alpha(b)$ is the area-weighted mean absorption coefficient in octave band $b$. In practice, in-situ measurements of RIRs and volumes of users’ local acoustic spaces are typically difficult to be carried out \cite{Jin16}. Alternatively, an attractive option is to blindly estimate room acoustic parameters from audio recordings using microphones. The 2015 ACE challenge \cite{Eaton16} sets the bar for the blind estimation of \rts\ and DRR from noisy speech sequences. Meanwhile, room volume estimation has long been formulated as a classification problem \cite{Moore14,Peters12}. With recent advancements in DNNs, formulating the blind room volume estimation as a regression problem by taking advantage of convolutional neural network (CNN) models in conjunction with time-frequency representations has become increasingly relevant. Genovese et al. \cite{Genovese19} deployed a CNN trained using both simulated as well as real RIRs, and results show that it can estimate a broad range of volumes within a factor of 2 on real-measured data from the ACE challenge \cite{Eaton16}. Similar CNN-based systems were proposed to blindly estimate room acoustic parameters from single-channel \cite{Gamper18,Bryan20,Gotz22} or multi-channel \cite{Srivastava21} speech signals and demonstrated promising results in terms of both estimation accuracy and robustness to temporal variations in dynamic acoustic environments. In addition to the log-energy calculation of spectro-temporal features that prior works generally relied on, Ick et al. \cite{Ick23} introduced a series of phase-related features and demonstrated clear improvements in the context of reverberation fingerprint estimation on unseen real-world rooms.

CNNs are widely considered in the aforementioned approaches due to their suitability for learning two-dimensional time-frequency signal patterns for end-to-end modelling. In order to better capture long-range global context, CNN-attention hybrid models that concatenate CNN with a self-attention mechanism have achieved cutting-edge results for numerous tasks such as acoustic event classification \cite{Kong20,Gong21} and other audio pattern recognition topics \cite{Li2018,Rybakov20}. Gong et al. \cite{Gong21ast} took one step further and devised purely attention-based models for audio classification. The devised Audio Spectrogram Transformer (AST) was assessed on various audio classification benchmarks with new state-of-the-art results, which shows that CNNs are not indispensable in this context.


Inspired by the work in \cite{Gong21ast}, in this work we propose a convolution-free, purely attention-based model to estimate geometric room volume blindly from single-channel noisy speech signals. To the authors' best knowledge, this is the first attention-based system in the area of blind acoustic room parameter estimation. The proposed system takes Gammatone magnitude spectral coefficients as well as the low-frequency phase spectrogram as inputs and captures long-range global context, even in the lowest layers. In addition, the system performance is further boosted by applying transfer learning of knowledge from an ImageNet-pretrained transformer model. A corpus of RIRs that consists of publicly available RIRs, synthesized RIRs and in-house measurements of real-world rooms is formulated with the aim of training and testing the proposed method. Experimental results show that the proposed model significantly outperforms CNN-based blind volume estimation systems on unseen real-world rooms using single-channel recordings.

\section{System Methodology}
In this section, we provide a detailed description of the proposed  attention-based system for blind room volume estimation. We start with an explanation of how we formulate input features that leverage both magnitude spectral and phase-related information. We then outline the design of the convolution-free model architecture. Finally, we highlight the use of transfer learning to enhance model performance with limited training datasets. Note that the geometric room volume is the main focus of this study. However, the proposed system can be readily extended for blindly estimating other room acoustic parameters (e.g. \rts, and total surface area). 

\subsection{Featurization}
Before being fed into the neural network, noisy speech signals go through a featurization process to obtain a two-dimensional time-frequency representation.  Various extracted features are integrated into the feature block for model training, aiming to effectively capture information about the acoustic space.

Similar to prior literature \cite{Genovese19, Gamper18},  the Gammatone ERB filterbank is selected for generating time-frequency representation as it leads to low model complexity while preserving signal information that is relevant to this problem. Specifically, the Gammatone filterbank consists of 20 bands covering the range from 50Hz to 2000Hz. We compute the audio by continuing with a Hann window of 64 samples and a hop size of 32 samples, followed by convolution with the filterbank. This convolution generates a spectral feature ($20 \times 1997$). Additionally, the phase information of the audio is retained. The phase angles computed for each time-frequency bin can be used to generate a phase feature, and it is then truncated to only include the bands corresponding to frequencies below 500 Hz (i.e. $5 \times 1997$) as lower frequency behavior generally carries more information corresponding to the room volume \cite{Kuttruff16}. Furthermore, the first-order derivative of phase coefficients along the frequency axis is also concatenated (i.e. $5 \times 1997$). The configuration of this feature set aligns with the ``+\textit{Phase}" model outlined in \cite{Ick23}, which is shown to outperform other methods that rely solely on magnitude-based spectral features. 

Overall, The proposed feature block has dimensions of $30 \times 1997$, where 30 represents the feature dimension  ${F}$, and 1997 represents the time dimension ${T}$.

\subsection{Model Architecture}
While the attention-based model demonstrates impressive performance in audio classification tasks, its application in other domains especially regression-related problems remains unexplored. In this section, we propose a purely attention-based model following the Audio Spectrogram Transformer work in \cite{Gong21ast} for conducting blind room volume estimation tasks.
\subsubsection{Audio Spectrogram Transformer}

\begin{figure}
    \centering
    \includegraphics[height=6.86cm]{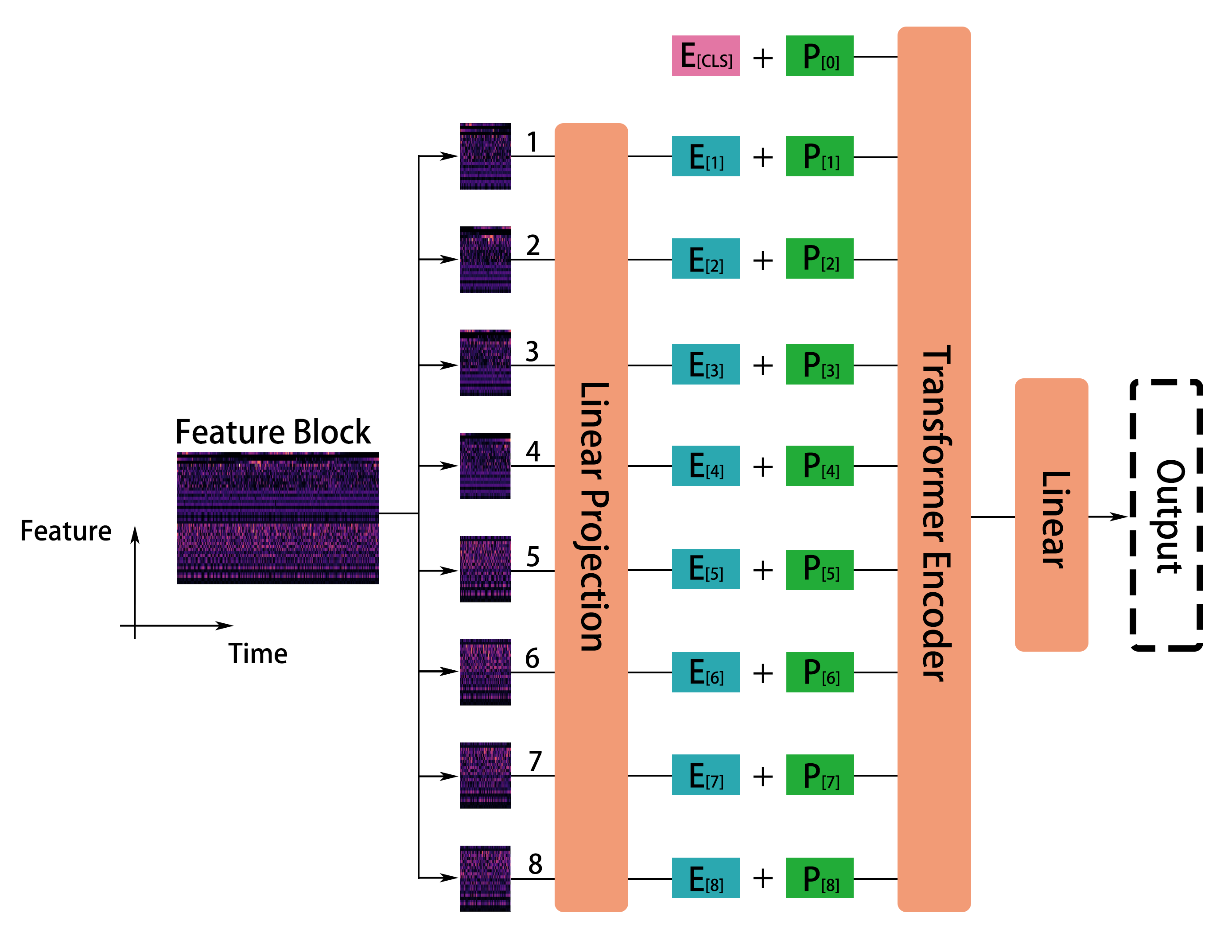}
     \label{fig:ast}
       \vspace{-6.5mm}
    \caption{Proposed system architecture with the featurization process.}
    \vspace{-4mm}
\end{figure}


In the proposed system, as shown in Fig. 1,  in order to better leverage the local information of the audio, feature blocks are divided into ${P}$ patches, each patch having a size of $16 \times 16$. During this division process, to maintain continuity in both the feature dimension and the time dimension, each patch overlaps with its surrounding patches by 6 feature dimensions and 6 time dimensions. Consequently, the number of patches ${P}$ is determined as 398, where ${P = \bigl\lceil \displaystyle \frac{F-16}{10} \bigr\rceil \bigl\lceil \displaystyle \frac{T-16}{10} \bigr\rceil}$. For further processing of these patches, a linear projection layer is introduced. This layer flattens each $16 \times 16$ patch into a one-dimensional patch embedding of size 768, referred to as the patch embedding layer.

Due to the fact that traditional transformer architectures do not directly process the sequential order of input sequences and these patches are not arranged in chronological order, trainable positional embeddings (which also have a dimension of 768) are incorporated into each patch embedding. This incorporation allows the model to grasp the spatial structure of the audio spectrogram and understand the positional relationships among different patches.

Similar to \cite{Gong21ast}, this paper also leverages a [CLS] token at the beginning of the sequence and feeds the feature sequence into the Transformer. In the proposed system, encoding and the feature extraction of the input sequence are achieved by utilizing only the encoder part of the original Transformer architecture \cite{AV17}. We adjust the input and output dimensions of the encoder. To be more precise, the input is a sequence formed by a feature block of size 30×1997 while the output is a single label used for volume prediction. The output of the whole Transformer is then used as the feature representation of the two-dimensional audio feature block, which is subsequently mapped to labels used for volume estimation through a linear layer with sigmoid activation.

\subsubsection{ImageNet Pretraining}
Compared to methods based on CNN architecture, one disadvantage of Transformer methods lies in their increased demand for training data \cite{AD21}. One of the main challenges in blind room volume estimation problems is due to insufficient data as publicly available RIR datasets with properly labelled room volume groudtruth are highly limited. To alleviate this issue, we took the following two measures: 1) a synthetic RIR dataset based on the image-source model (which will be covered in Sec. 3.1), 2) transfer learning. 

More specifically, cross-modality transfer learning was applied to the proposed Transformer-based model. In this context, we leveraged a pretrained off-the-shelf Vision Transformer (ViT) model from ImageNet \cite{HT20} for application within the proposed method. Prior to the transfer, necessary adjustments were made to ensure ViT's compatibility with the proposed architecture. Firstly, the ViT model was pretrained on three-channel images, which was distinct from the single-channel feature blocks used in the proposed model. Therefore, we calculated the average of parameters across the three channels of the ViT model and then applied this averaged information. In addition, the so-called 'Cut and bi-linear interpolate' method \cite{Gong21ast} was used to adjust the input size and manage positional encoding. Lastly, to adapt ViT for the task of blind room volume estimation, the final classification layer of ViT was reinitialized.

\section{DATA GENERATION AND AUGMENTATION}
To address the challenging task of room volume estimation, neural networks require extensive data to train and validate. In this section, we devise a multi-stage audio generation process, utilizing six publicly available real-world RIR datasets and a synthetic dataset based on room simulation. In addition, a series of data augmentation techniques are introduced to enhance the generalizability of the model.
\subsection{RIR Dataset}
As shown in Fig. 2, six publicly available real-world RIR datasets recorded in 55 real rooms are considered to cover a wide range of realistic acoustic room parameters. Data collection predominantly took place in geometrically regular rooms, encompassing spaces such as elevator shafts, classrooms, auditoriums, seminar rooms, and more. These datasets include the ACE Challenge dataset \cite{Eaton16}, the Aachen Impulse Response (AIR) dataset \cite{Jeub09}, the Brno University of Technology Reverb Database (BUT ReverbDB)  \cite{Igor19}, the OpenAIR dataset \cite{murphy2010openair:}, the C4DM Dataset (C4DM) \cite{Stewart10} and the dechorate Dataset (dechorate) \cite{Carlo21}. To account for the natural gap of real-world acoustic spaces between the 12 $m^3$ to 7000 $m^3$  range of volumes, the in-house BJUT Reverb Dataset was collected. Specifically, we took RIR measurements at 11 distinct rooms within the campus of Beijing University of Technology. For each room, 5 RIRs corresponding to different source-receiver positions were recorded. All RIRs are resampled to the sampling rate of 16kHz.
\begin{figure}
    \centering
    \includegraphics[height=3.7cm]{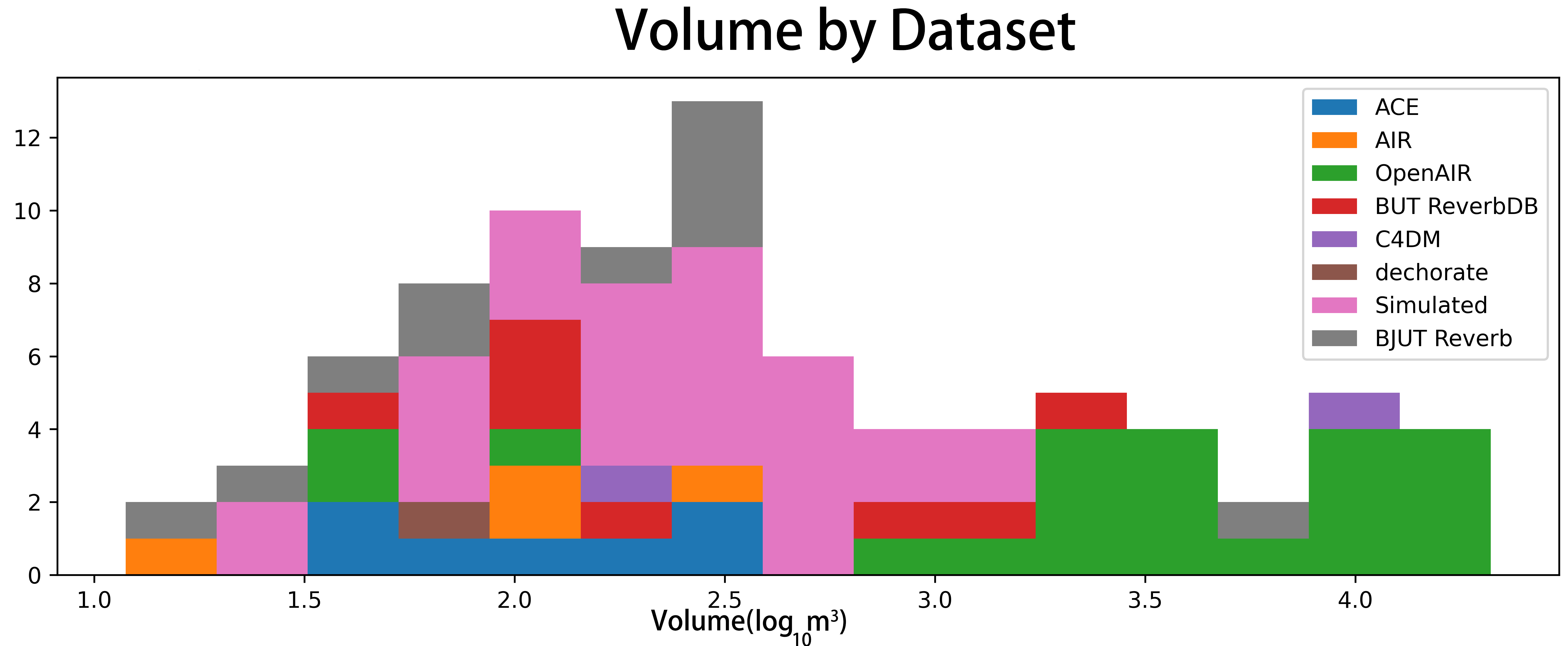}
    \label{fig:rirs}
     \vspace{-6mm}
    \caption{The histogram illustrates the distribution of RIRs across various datasets based on their labelled volumes}
    \vspace{-3mm}
\end{figure}

 Moreover, we supplemented the real-world data with an additional 30 simulated RIRs based on virtual rooms of various geometries. The purpose was to augment the dataset within less common room volume ranges, thus bringing the total volume distribution closer to a normal distribution. To generate the synthetic dataset, we utilized the pyroomacoustics \cite{Scheibler17} package that employs the image-source model to simulate RIRs for rooms with specific volumes.

\subsection{Audio preprocessing}
By utilizing the RIR dataset with labelled volumes, convolution was applied to 4-second clean speech signals recorded in anechoic chambers from the ACE dataset \cite{Eaton16}, resulting in reverberant speech sequences that were characterized by corresponding RIRs. To enhance the model's adaptability to noise across various types and signal-to-noise ratios (SNR), white noise and babble noise were added. Each type of noise was applied at the following SNR levels: [+30, +20, +10, +0] dB. This formulated \emph{Dataset I}, which  was then divided into train, test, and validation sets in a 6-2-2 ratio (as shown in Table 1). Note that for the test set, only RIRs recorded in real-world environments were selected to assess the model's estimation performance on unseen non-simulated rooms.

Moreover, for the sake of enhancing the networks' generalizability in unknown rooms and noisy environments, we adapted the widely-used SpecAugment \cite{Park} augmentation scheme and added it to our data generation pipeline. Specifically, reverberant speech signals without noise addition were considered in the training set. Subsequently, these audio were transformed into log mel spectrograms, upon which time masking, frequency masking and time warping were applied, as illustrated in Fig. 3. Then, masked mel spectrograms were converted back into time-domain signals. Finally, the resulting 4800 speech sequences with masking effects were added to the original training set for neural network training, which is denoted as \emph{Dataset II}.  

\begin{figure}
    \centering
    \includegraphics[height=4.15cm]{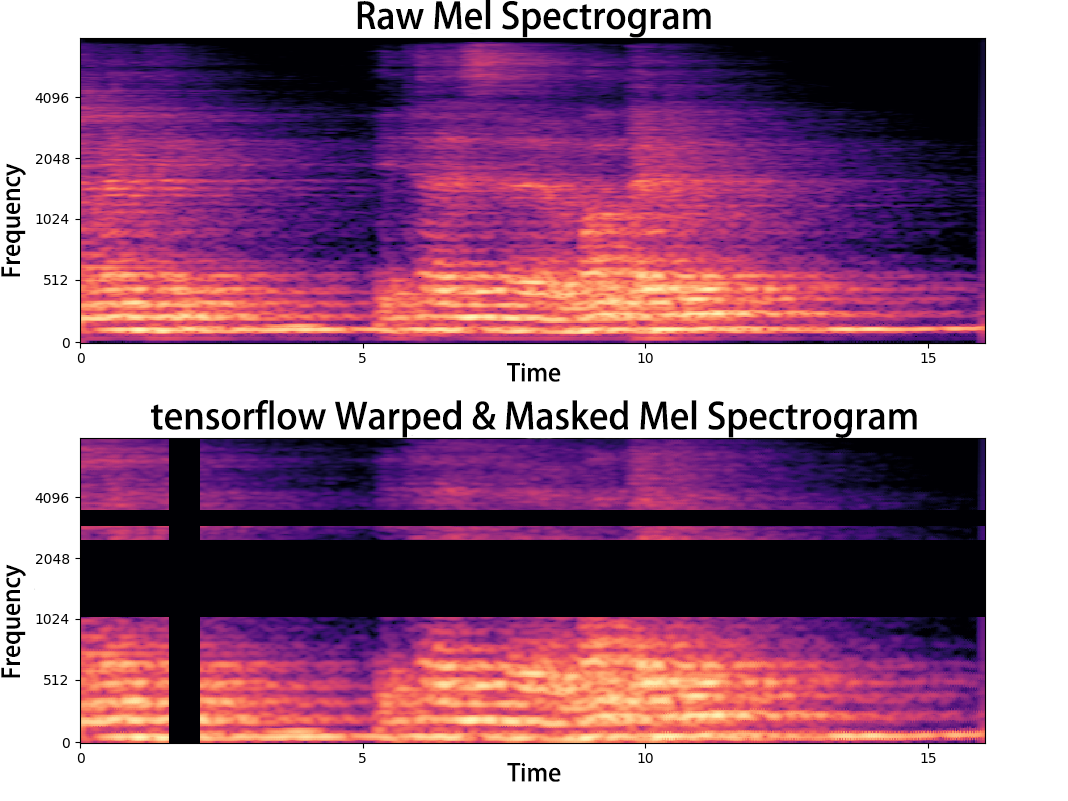}
    \label{fig:specs}
     \vspace{-3mm}
    \caption{Augmentation schemes applied to reverberant speech signals.}
    \vspace{-1mm}
\end{figure}

\begin{table}
\vspace{1mm}
\begin{tabular}{c|cccc}
\centering
Data  &  {$\#$} of & {$\#$} of   & Real & Simulated\\
Spilt & \emph{Dataset I}  & \emph{Dataset II} & Rooms &  Rooms\\
\hline
Train & 19200  & 24000 & 34 & 18 \\
Validation & 6400  & 6400 &  21 & 12  \\
Test & 6400 & 6400 & 21 & 0 \\
\end{tabular}
\vspace{-1mm}
\caption{Summary of Data Splits for \emph{Datasets I} \& \emph{II}}
\vspace{-5mm}
\label{table:sum}
\end{table}

\begin{table*}[h]
\centering
\begin{tabular}{c|c c c p{1cm}|c c c p{1cm}}
\hline
\multirow{2}{*}{Method} & \multicolumn{4}{c|}{\emph{Dataset I}} & \multicolumn{4}{c}{\emph{Dataset II}} \\

\cline{2-9}
&MSE &MAE & $\rho$ & \textit{MM} & MSE &MAE & $\rho$ & \textit{MM} \\
\hline
CNN \cite{Ick23} & 0.3863 & 0.4837 & 0.6984 & 3.0532 & 0.3154 & 0.4136 & 0.7678 & 2.5921 \\
\hline
Proposed method& 0.2650 & 0.3432 & 0.8077 & 2.2039 & 0.1981 & 0.2884 & 0.8580 & 1.9427 \\
\hline
\textbf{Proposed method}  & \multirow{2}{*}{0.2157} & \multirow{2}{*}{0.3111} & \multirow{2}{*}{0.8529} & \multirow{2}{*}{{2.0470}} & \multirow{2}{*}{\textbf{0.1541}} &\multirow{2}{*}{\textbf{0.2423}} & \multirow{2}{*}{\textbf{0.8929}} & \multirow{2}{*}{\textbf{1.7470}} \\
\textbf{w/pretrain}  & & & & & &\\

\hline
\end{tabular}
\vspace{-1mm}
\caption*{\textbf{Table 3}: Performance comparison of different models with and without the application of SpecAugment.}
\vspace{-1mm}
\label{per2}
\end{table*}

\section{Experimental Results}
In this section, we evaluate the effectiveness of the proposed attention-based method and compare it to the state-of-the-art approach in the realm of single-channel blind room volume estimation. First, the experimental design and setup of training sessions are introduced. Second, we present results that demonstrate the estimation results of the considered systems in two different tracks.
\subsection{Experimental Design}
To assess the performance of our proposed approach, we compared it with the \emph{+Phase} model in \cite{Ick23} that leverages a CNN-based architecture, as well as phase-related feature sets. This CNN model consists of six convolutional layers, each followed by an average pooling layer. Following the convolutional blocks are a dropout layer and a fully connected layer mapping to the output dimension, forming a complete feedforward convolutional neural network. This method is considered as the state-to-the-art in terms of single-channel blind room volume estimation.

We evaluated our data on the base-10 logarithm of the volume, which implies that the estimation error would be related to its order of magnitude. A log-10 estimate is more appropriate than a linear one due to the significantly large range of room volumes in the test set as shown in Fig. 2. The following four metrics were considered in this evaluation: mean squared error (MSE), mean absolute error (MAE), the Pearson correlation coefficient ($\rho$) between predicted and target values, and MeanMult (\textit{MM}). \textit{MM} is the mean absolute logarithm of the ratio between the estimated room volume $\hat{V}_n$ and the ground truth $V_n$:

\begin{equation}
    \textit{MM} = e^{\frac{1}{N} \sum_{n=1}^{N} | \ln\left(\frac{\hat{V}_n}{V_n}\right) |}
\end{equation}
 where $N$ is the number of samples. This metric provides an overview of the average error in ratios between estimated and target parameters.

During the model training phase, MSE was used as the loss function and the Adam optimizer from PyTorch was deployed for optimization. The CNN model and the proposed method were trained for 1000 epochs and 150 epochs, respectively. This decision was made as good convergence behaviors were already observed for both training and validation at these epoch numbers. L2 regularization was applied to mitigate potential over-fitting. Additionally, an adaptive learning rate strategy was adopted to ensure efficient convergence during training. If the model failed to demonstrate improvement on the validation set for ten consecutive epochs, early stopping criteria were applied to halt the training process. For comparison test purposes, we switched between \emph{Dataset I} and \emph{Dataset II}, as well as whether or not to use the pretrained model from ImageNet, while maintaining same hyperparameters between the attention-based and CNN-based model to ensure uniformity in model configurations.
\vspace{-1mm}
\subsection{Results}
\subsubsection{Base Systems}

We started with the experiment of comparing the base version of our proposed method with the CNN-based baseline system \cite{Ick23}. The goal of this experiment is to observe if we can achieve similar estimation performance by simply replacing CNN with a purely attention-based model. We trained both the CNN model and the proposed method (without ImageNet pretraining) on \emph{Dataset I} separately, feeding them with the same feature set as outlined in Section 2.1. Results of these two models are presented in Table \ref{table:per1}.

It can be seen that the proposed model significantly outperforms the CNN model in terms of prediction accuracy, relationship with ground truth values and predictive capability. This indicates that neural networks purely based on attention are sufficient (or even more superior than CNN models) to accurately learn the relationship between indoor acoustic characteristics and room volumes, even with the low-layer network configuration and a relatively small number of training epochs.
\vspace{-1mm}
\subsubsection{Enhanced Systems With Pretraining}
\begin{table}
\centering
\begin{tabular}{c|cccc}
Method & MSE & MAE & $\rho$ & \textit{MM} \\
\hline
CNN \cite{Ick23} & 0.3863  & 0.4837 & 0.6984 & 3.0532 \\
\textbf{Proposed method} & \textbf{0.2650}  & \textbf{0.3432} &  \textbf{0.8077} & \textbf{2.2039} \\
\end{tabular}
\caption{Comparison between the CNN-based system \cite{Ick23} and the base version of the proposed method.}
\vspace{-7mm}
\label{table:per1}
\end{table}
To further investigate the impact of ImageNet pretraining on the performance of the proposed method, we introduced the ``Proposed method w/ Pretrain" model. We conducted separate training sessions for the CNN model, the Proposed method, and the ``Proposed method w/ Pretrain" model on \emph{Dataset I}. Additionally, we also incorporated the SpecAugment data augmentation method into the three different models. Specifically, all three models were retrained on \emph{Dataset II} to investigate its impact on model performance. The results of the above experiments are listed in Table 3.

With \emph{Dataset I}, the deployment of the ImageNet pretraining elevated the proposed method's performance to a new level, yielding a significantly improved room volume estimation accuracy. With the application of the SpecAugment method, \emph{Dataset II} facilitated to further enhance system performance for all three models. Particularly, the augmentation effect was more prominent in the proposed method w/ Pretrain, confirming the effectiveness of SpecAugment in terms of alleviating overfitting and enhancing models' generalizability. As a more illustrative example, the best-performing system, i.e. ``Proposed method w/ Pretrain" model with \emph{Dataset II}, resulted in a median and mean absolute error of only 155 $m^3$ and 1219 $m^3$ in linear scale respectively, given that the range of test set room volumes was [12, 21000] $m^3$. In contrast, the median and mean absolute error of the CNN-based system with \emph{Dataset II} was 353 $m^3$ and 1919 $m^3$, respectively. 

Fig. 4 demonstrates the confusion matrices for these two systems, with the x-axis and y-axis representing log-10 indices for volume sizes. It can be clearly seen that the ``Proposed method w/Pretrain" model is consistently well-distributed around the ground truth across the tested range while the CNN-based method diverges. This indicates that our proposed attention-based model captures the representation of the room volume regression problem through the effective training process and more importantly generalizes the learned patterns to unseen real-world rooms.

\begin{figure}
    \centering
    \includegraphics[height=4.10cm]{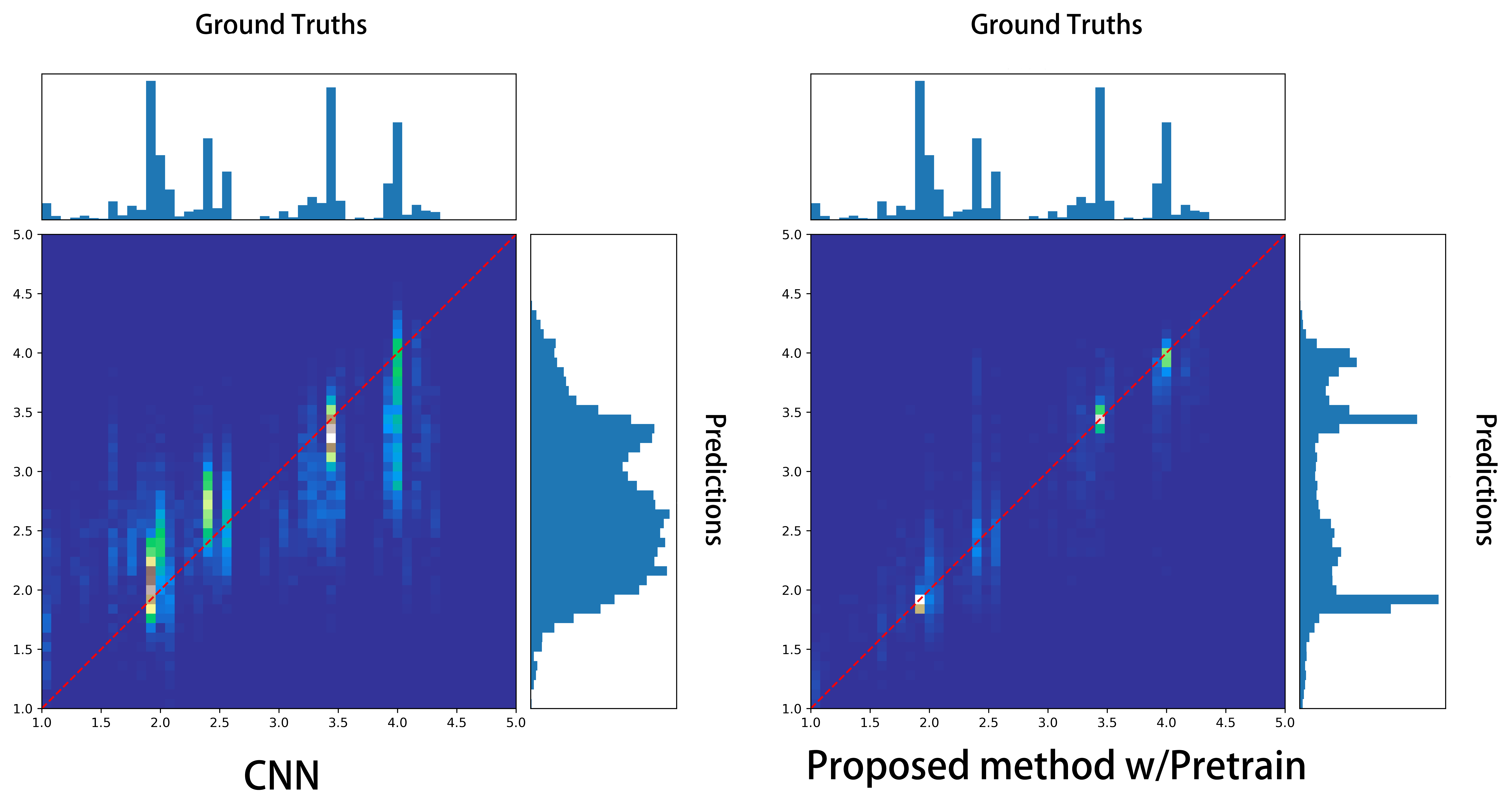}
    \label{fig:hots}
    \vspace{-3mm}
    \caption{Confusion matrices for the CNN model and the ``Proposed method w/Pretrain" model trained on \emph{Dataset II}. The dashed red line indicates a perfect prediction.}
    \vspace{-3mm}
\end{figure}

\section{Conclusion and Future work}
In  this study, we aim to explore the feasibility of applying a Transformer-based model in the blind room volume estimation task and to benchmark its performance using different training strategies. Experimental results based on unseen real-world rooms with realistic noise settings confirm that the proposed method exhibits more superior performance compared to traditional CNN-based methods, indicating that a neural network
purely based on attention is sufficient to obtain high performance in audio-related regression problems. Future work will investigate the flexibility and robustness of the proposed system in terms of variable-length audio inputs.

\bibliographystyle{IEEEbib}
\bibliography{refs}

\end{document}